\documentstyle[12pt,a4]{article} 
\input epsf                                            
\begin{document}
\vspace*{-2cm}
\hfill{\bf IReS 98-05}
\vspace{1.5cm}

\begin{center}
{\Large\bf Comparison of \mbox{\boldmath $\eta$} and \mbox{\boldmath $\eta'$}
production in the}\\[2ex]
{\Large\bf \mbox{\boldmath $p\,p \to p\,p\,\eta(\eta')$} reactions
near threshold}\\[7ex] 
F.~Hibou$^1$, O.~Bing$^1$, M.~Boivin$^2$, P.~Courtat$^3$, G.~F\"{a}ldt$^4$, 
R.~Gacougnolle$^3$, Y.~Le~Bornec$^3$, J.M.~Martin$^3$, A.~Moalem$^5$, 
F.~Plouin$^{2}$, A.~Taleb$^1$, B.~Tatischeff$^3$, C.~Wilkin$^6$, 
N.~Willis$^3$, R.~Wurzinger$^{2,3}$.\\[5ex]

$^1$ Institut de Recherches Subatomiques, CNRS--IN2P3 / Universit\'e
Louis Pasteur, B.P.28, F--67037 Strasbourg Cedex~2, France\\[1ex]
$^2$ Laboratoire National SATURNE, F--91191 Gif-sur-Yvette Cedex, France\\[1ex]
$^{3}$ Institut de Physique Nucl\'eaire, IN2P3--CNRS,
F--91406 Orsay Cedex, France\\[1ex]
$^{4}$ Nuclear Physics Division, Uppsala University, Box 535,
S-751~21 Uppsala, Sweden\\[1ex]
$^{5}$ Physics Department, Ben Gurion University, 84105 Beer Sheva,
Israel\\[1ex]
$^{6}$ University College London, London WC1E 6BT, United Kingdom\\[10ex]
\end{center}

\begin{abstract}
The total cross section of the $p\,p \to p\,p\,\eta'$ reaction has
been measured at two energies near threshold by detecting the final 
protons in a magnetic
spectrometer. The values obtained are about a factor of 70 less than for the
corresponding $\eta$ production, in good agreement with the predictions of a
one-pion-exchange model.\\[3ex]
\end{abstract}

\noindent
{\em PACS:} 13.60.Le, 13.75.Cs, 14.40.Cs\\
\noindent
{\em Keywords:} $\eta$, $\eta'$ mesons, mass, production threshold.\\[6ex]

\newpage
The production of $\eta$ mesons in proton-proton collisions near threshold has
been measured in recent years using three different techniques. By detecting
both final protons in a magnetic spectrometer, the SPESIII group at Saturne
\cite{Bergdolt,Taleb} observed the $\eta$ meson as a missing mass peak. 
The PINOT group \cite{PINOT}, also working at Saturne, instead identified 
the $\eta$ directly through its decay into two photons. Up to at least the 
threshold for $\eta\pi$ production this provides a clean measurement of 
the $pp\to pp\eta$ total cross section. The CELSIUS group used a combination 
of the two techniques when measuring the $\gamma\gamma$ decay in coincidence 
with the directions (but not the energies) of two final charged particles 
\cite{Calen}. Each method has its own advantages but it is very reassuring 
that the total cross sections from the three groups, shown in fig.~1
in terms of the centre-of-mass excess energy $Q$, present such a consistent 
picture. In the present letter we report on the first measurement of $\eta'$
production in proton-proton collisions near threshold using the magnetic
spectrometer technique, and compare the resulting cross sections with those 
obtained for the $\eta$.

The experiment was carried out at the Laboratoire National SATURNE
(LNS) using the SPESIII spectrometer under experimental conditions which were 
similar to those reported in ref.~\cite{Bergdolt} and so only essential 
features are reported here. SPESIII is a large acceptance magnetic 
spectrometer which is well suited to the study of meson production
involving both two and three-body final states near threshold.
The detection system consists of three multiwire drift chambers, the first 
situated on the focal plane, and four planes of scintillators, comprising in 
total 70 counters. Under its standard working conditions the spectrometer can 
analyse particles with momenta p/Z lying between 600 and 1400~MeV/c with a 
resolution $\delta p/p$ in the range $(0.5 - 1.0)\times 10^{-3}$. 

The maximum opening angles in both the horizontal and vertical directions 
are $\pm 60$~mr and the effective solid angle acceptance is 
$\Delta\Omega \approx 10^{-2}$~sr. 
The momentum resolution can be improved by using
one of a series of collimators, though at the expense of reducing the
solid angle, and this was done for most of the $\eta$ measurements
reported in ref.~\cite{Bergdolt}. In view of the smaller cross section
expected for the $\eta'$, the collimators were here withdrawn to obtain the
maximum solid angle. In such conditions the spectrometer and its detection 
system covered about 85\% and 50\% of the total $pp\eta'$ phase space at the 
two beam energies studied, which were nominally 2416 and 2430~MeV. The liquid 
hydrogen target had a thickness of 270~mg/cm$^2$. The control of the beam 
intensity and measurement of the number of incident protons was carried out 
with the help of two scintillator telescopes viewing the target and an 
ionisation chamber placed in the beam downstream of the target. The absolute 
calibration of these counters was carried out using the standard carbon 
activation technique \cite{Banaigs}.

The missing mass spectra of the $pp\to pp X$ reaction shown in figs.\ 2(a1)
and (a2) correspond to measurements a little above the $\eta'$ threshold
at $T_p =$~2416 and 2430~MeV respectively. The clear rise near the end of 
these spectra contains $\eta'$ events but superimposed on a continuum of
multipion production and this continuum must be subtracted in order to
determine the 
numbers of $\eta'$ produced. This was carried out using analogous spectra
recorded just below threshold at $T_p =$~2400~MeV. The measured momenta
$p_1$ and $p_2$ for each event were increased by an amount $\Delta p$ and
the missing mass was evaluated as if the beam energy were either 2416 or 
2430~MeV. The values of $\Delta p$ were fixed by demanding that the range
of proton momenta near the highest missing masses was the same above and
below threshold. At the endpoint of the spectra, $\Delta p = m_p (\beta'
\gamma'-\beta\gamma)$, where $\beta'$ and $\beta$ are the incident proton
speeds above and below threshold. The predicted backgrounds shown in
figs.~2(b1) and 2(b2) still manifest some enhancement near the ends of the 
spectra arising from the proton-proton final state interaction (fsi). 
The $\eta'$ peaks were obtained by subtracting these background
spectra from those of the $pp\to pp X$ measured above the $\eta'$ production 
threshold. The shapes and widths of the $\eta'$ peaks in the resultant
$pp\to pp\eta'$ spectra, shown in figs.~2(c1) and 2(c2), are in good agreement
with the computer simulations discussed later.

In view of the rapid variations of the three-body cross section and the
spectrometer acceptance near threshold, it is vital to determine the 
excess energy $Q$ in the centre-of-mass (cm) system with
precision. Now the nominal beam energy calculated
using the parameters of the SATURNE accelerator is in general slightly greater
than the true value. Several previous detailed studies 
\cite{Taleb,Willis,Plouin} show that the difference between
the real and nominal energies does not depend upon the energy or particle
type but rather on the tuning of the accelerator and on the beam extraction.
>From these measurements we determine an energy correction of
$dT = 1.1\pm 1.0$~MeV which changes a nominal beam energy of 2416~MeV to a
`true' value of $T_p= 2414.9$~MeV. Taken together with an $\eta'$ mass
of $m_{\eta'}= (957.77\pm 0.14)$~MeV/c$^2$ \cite{PDG}, this leads to a value 
of $Q=(3.5\pm 0.4)$~MeV.

It is however possible to determine the value of $Q$ directly from the
experimental data, as has been done previously at SPESIII for $\eta$ 
production \cite{Bergdolt,Willis}. Near threshold the width of the proton 
momentum spectrum is a sensitive function of $Q$ which depends but weakly upon 
the final masses and the absolute calibrations of the beam and spectrometer.
Fig.~3(a) shows such a spectrum for $\eta'$ production, obtained by selecting 
those events in figs.~2(a1) and 2(b1) in the range 956~MeV/c$^2 \leq
M_x\leq$~962~MeV/c$^2$ and performing a subtraction, to be compared with
a simulation performed assuming that $Q=4.3$~MeV. From the variation of
$\chi^2$ for such fits shown in fig.~3(b), we deduce that $Q=4.3\pm 0.9$~MeV,
corresponding to a value of $m_{\eta'}=(957.0\pm 0.9)$~MeV/c$^2$ for a beam
energy of 2414.9~MeV. This is in good agreement with the World
compilation of the $\eta'$ mass \cite{PDG}. The value of the mid-target value 
of $Q$ quoted in Table~1 is the mean of the two types of determination. 
Adjustments in the Saturne beam energy of a few MeV can be accomplished 
with very high precision and the value of $Q$ quoted for this in the table 
is based upon the change in nominal beam energy.

These values of $Q$ were entered into a calculation of the acceptances
which was carried out with a very refined simulation program taking into
account all the details of the experimental set-up as well as the algorithms
used for treating the data. This program, already used for the first
determination of the $pp\to pp\eta$ at SPESIII \cite{Bergdolt}, has
benefited from numerous upgrades, especially in regards of the effects 
of the proton-proton final state interaction. Thanks to such 
improvements the mass of the $\eta$ determined at SPESIII, 
$m_{\eta} = (547.65\pm 0.18)$~MeV/c$^2$ \cite{Taleb}, is in good agreement
with the PDG average of $m_{\eta} = (547.45\pm 0.19)$~MeV/c$^2$ 
\cite{PDG}. Updated values of the $pp\to pp\eta$ total cross section of
ref.\cite{Bergdolt} are given in table~1, along with the first measurements
of the $pp\to pp\eta'$ reaction, and also shown in fig.~1. The quoted errors 
take account of all the uncertainties both statistical and systematic,  
including an overall normalisation error of 10\%. It should be noted that at 
$Q=3.7$~MeV the acceptance is insensitive to small changes in the value of $Q$ 
and this will also be true for the deduced value of $\sigma_{T}$. On the 
other hand, a change of $\pm 0.7$~MeV at $Q=8.3$~MeV would result in a 
variation of $\pm 3.5$~nb in the quoted cross section. Theoretical models
are not currently sensitive to such small uncertainties.

In a one-pion-exchange model, a $\pi^0$ is emitted by one proton and 
converts to
an $\eta$ on the second, leaving the two final protons to be subject to the
strong S-wave final state interaction \cite{GW}. If the initial $pp$ distortion
is neglected, it is straightforward to show in such a model that the S-wave 
contribution to the total $pp\to pp\eta$ cross section near threshold should
have the form
\begin{equation}
\label{Eq:1}
\sigma_T(pp\to pp\eta) = A\:
\frac{(m_p+m_{\eta})^2}{(2m_p+m_{\eta})^{5/2}}\
\frac{\sqrt{m_{\eta}}}{(m_pm_{\eta}+m_{\pi}^2)^2}\: |f(\pi^0p\to p\,\eta)|^2\:
F(Q)\:,
\end{equation}
with an analogous formula for $\eta'$ production. Here $m_p$, $m_{\eta}$ and 
$m_{\pi}$ are respectively the $p$, $\eta$ and $\pi$ masses. Despite the
dominance of the N$^*(1535)$ resonance, the spin-average of the square of the 
$\pi^0p\to p\,\eta$ amplitude, $|f|^2$, varies slowly over the limited range 
of $Q$ measured in this experiment. It is therefore sufficient to take its
value at threshold.

In the short-range limit, which is valid for heavy-meson production, the 
effect of the proton-proton final state interaction can be modelled by the 
function \cite{FW} 
\begin{equation}
F(Q)= \epsilon\left(\frac{Q}{\epsilon}\right)^2
\left(1+\sqrt{1+Q/\epsilon}\,\right)^{\!-2}\:,
\end{equation}
where, including Coulomb distortion, $\epsilon\approx 0.45$~MeV. Such a 
function describes accurately the near-threshold energy dependence of the 
$pp\to pp\pi^0$ total cross section down to $Q\approx 1$~MeV, when 
explicit Coulomb suppression becomes important \cite{FW}.

The overall factor $A$ depends upon the $\pi pp$ coupling constant and the 
proton mass {\it etc.}, but is independent of parameters referring to the $\eta$
meson, so that its value should be the same for both $\eta$ and $\eta'$ 
production.

The dashed curve in fig.~1 shows the prediction of eq.(\ref{Eq:1}) when the
value of $A$ is adjusted to fit the lowest energy points of $\eta$ 
production. The deviations from this curve of up to a factor of two at higher 
values of $Q$ have been taken as evidence for an attractive $\eta pp$ final 
state interaction \cite{Calen}. By using the same value of $A$, combined with experimental values for 
the threshold values of $|f(\pi^0p\to p\,\eta)|^2 =(365\pm 30)\,\mu$b/sr and
$|f(\pi^0p\to p\, \eta')|^2 =(10\pm 1)\,\mu$b/sr \cite{Binnie}, it is possible 
to make absolute predictions of the $pp\to pp\eta'$ total cross section, and 
the resulting solid line shown in fig.~1, is only about 30\% below our two
experimental points and in good agreement with the trend in the data. Though
the predicted energy dependence is only weakly model dependent, more
accurate estimates of the normalisation would require further information
on the contributions of $\rho$ and other heavy-meson exchange diagrams
\cite{GW}.

In summary, we have made a first measurement of the $pp\to pp\eta'$ total
cross section near threshold, determining the mass of the $\eta'$ to better
than 0.1\%. The relative production rates of the $\eta'$ and $\eta$ mesons 
are broadly consistent with expectations based on a meson exchange
model. Since the preparation of this letter, we have been informed
that measurements by the COSY-11 group closer to threshold have been
submitted for publication \cite{COSY11}. It should be noted that there
is satisfactory agreement in normalisation where the two experiments
meet around $Q= 4$~MeV. \\

We wish to thank F.~Brochard for much help and advice.
The SATURNE machine was closed down on the night of 1/2 December 1997. We 
should like to express our gratitude to the accelerator crew and support staff 
for providing us with excellent conditions in which to work on this and 
previous experiments.
\vspace{1cm}

%\newpage

\newpage
\begin{table}[t]                                                
\centering                                                       
\label{table1}                                                   
\caption{Total cross sections for $\eta$ and $\eta'$ production in 
proton-proton collisions near threshold measured with SPESIII. The excess 
energy $Q$ is the mean 
kinetic energy in the centre-of-mass system of the final state, taking into
account the energy losses of the beam in the target. The inclusion of
proton-proton final state interactions in the analysis led to a 
significant reduction of the $\eta$ cross section at 16~MeV, as compared to 
the value quoted in [1], and this might be reduced by 
up to a further 10\% by a strong $\eta$-proton fsi.}
\vspace{8ex}                                                     
\begin{tabular}{lll}\hline                                
&&\\                                                            
&$Q$ (MeV)&$\sigma_{T}$ (nb)\\
&&\\                                                            
\hline  
&&\\
$pp\to pp\eta$&$0.64\pm 0.25$&$\phantom{88}80\pm 15$\\
              &$2.65\pm 0.25$&$\phantom{8}730\pm 120$\\
              &$16.0\pm 0.6$&$2680\pm 540$\\
&&\\
\hline
&&\\
$pp\to pp\eta'$&$\phantom{8}3.7\pm 0.6$&$\,19.2\pm 2.7$\\
               &$\phantom{8}8.3\pm 0.7$&$\,43.6\pm 6.5$\\
&&\\
\hline                                                           
\end{tabular}                                                    
\vspace{1cm}
\end{table}

\clearpage

\begin{figure}
\begin{center}
\mbox{\epsfxsize=14cm \epsfbox{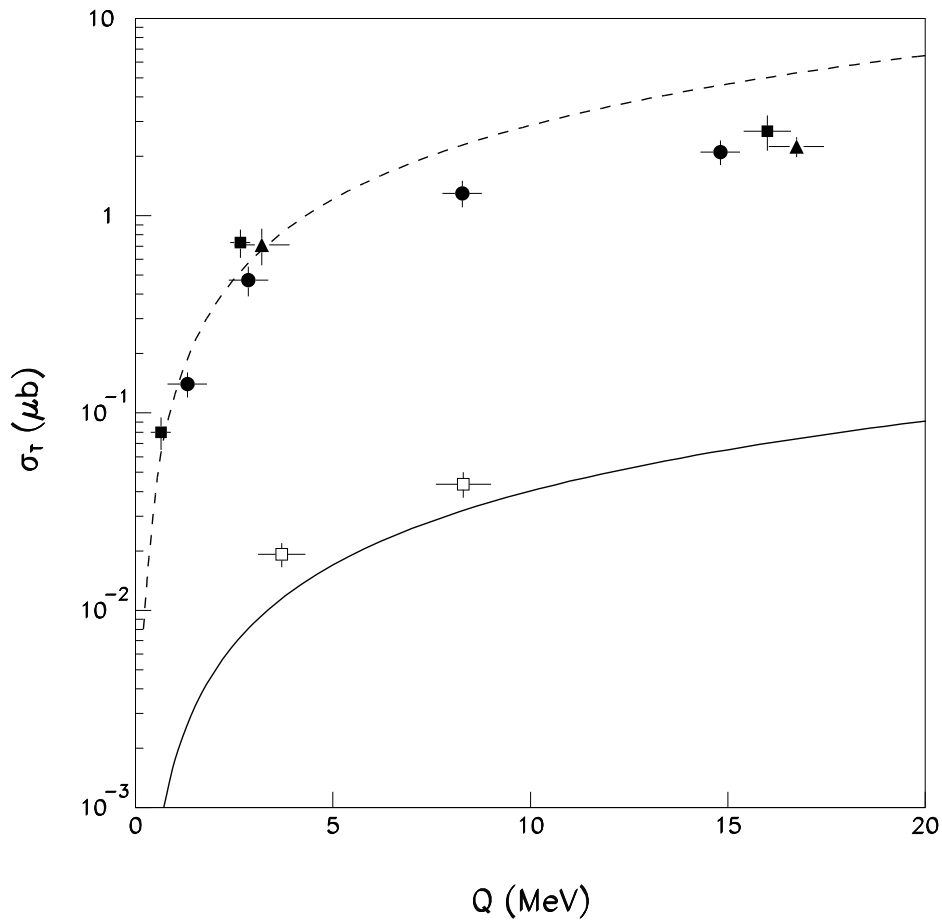}}
%\epsf{/users/tissier/sauveat/amoi98/hibfig1.eps,width=14cm}
%\mbox{\epsfxsize=14cm \epsfbox{/users/tissier/sauveat/amoi98/pubs/hibfig1.eps}}
\end{center}
\caption{Total cross sections for $pp\to pp\eta$ as a function of the mean
mid-target kinetic energy $Q$ in the final state measured with SPESIII (filled
squares) [1]. The CELSIUS points (circles) [4] and those of PINOT (triangles) 
[3] are subject to additional overall normalisation uncertainties of 
$\pm 20$\% and $\pm 15$\% respectively, whereas the $\pm 10$\% of the
SPESIII points are included in the figure. The PINOT energies are slightly 
uncertain and the mean of their two solutions is shown, corrected for energy 
loss in the target. The points are compared to the predictions of
eq.(1) (dashed curve), 
which reflects the proton-proton final state interaction folded with phase 
space. The arbitrary normalisation in  eq.(1) for the $\eta$ case fixes the 
absolute scale for the $pp\to pp\eta'$ predictions shown as the solid curve. 
This is in reasonable agreement with our two measured $\eta'$ points 
(open squares).}
\end{figure}

\begin{figure}
\begin{center}
\mbox{\epsfxsize=14cm \epsfbox{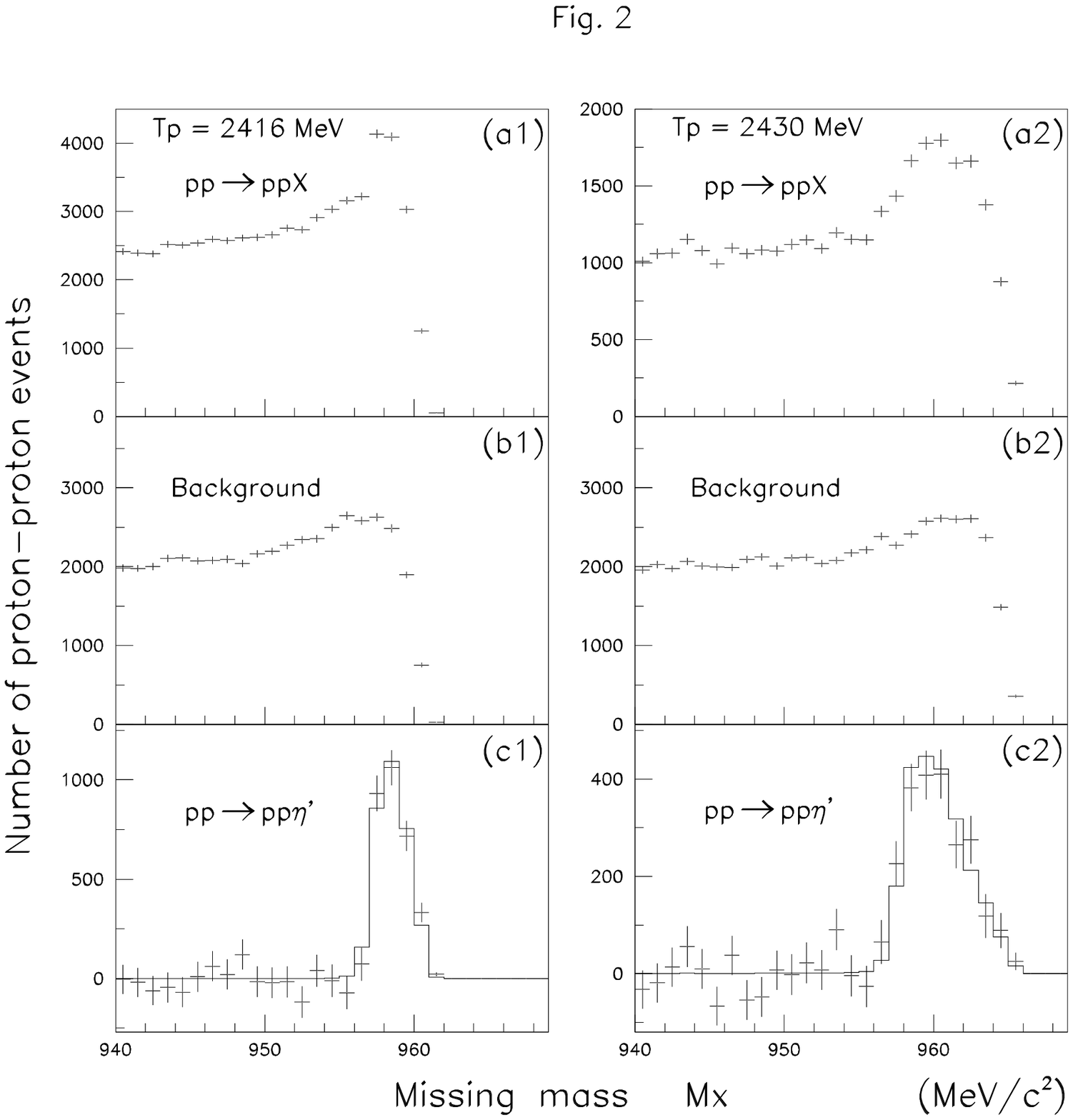}}
%\mbox{\epsfxsize=14cm \epsfbox{/users/tissier/sauveat/amoi98/pubs/hibfig2.eps}}
\end{center}
\caption{Missing mass spectra of the $pp\to ppX$ reaction at nominal beam
energies of (1) 2416~MeV, and (2) 2430~MeV. The observed spectra in
(a1) and (a2) are to be compared with background spectra in (b1) and (b2) 
obtained by scaling the 2400~MeV data as described in the text. The shapes 
and widths of the $\eta'$ peaks, obtained by subtraction and shown in 
(c1) and (c2), are in good agreement with the Monte Carlo simulations.}
\end{figure}

\begin{figure}
\begin{center}
\mbox{\epsfxsize=14cm \epsfbox{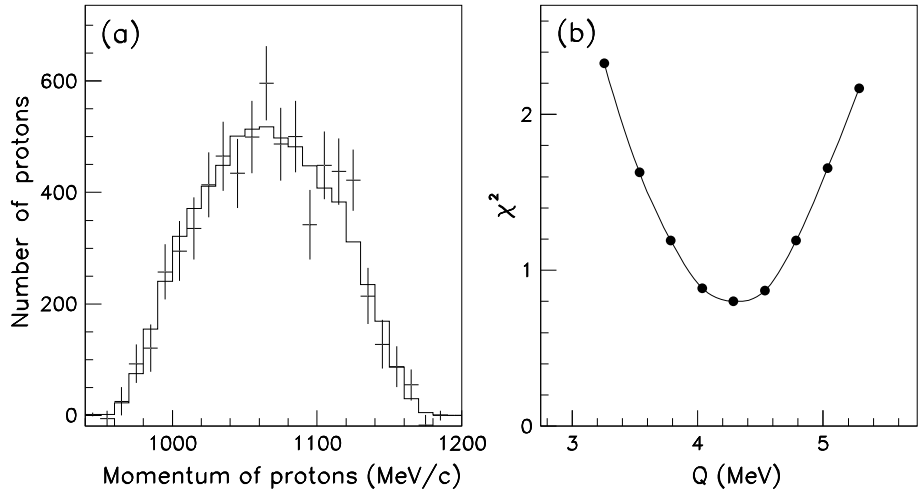}}
%\mbox{\epsfxsize=14cm \epsfbox{/users/tissier/sauveat/amoi98/pubs/hibfig3.eps}}
\end{center}
\caption{
(a) Proton momentum spectrum for events in the $\eta'$ peak in the
2416~MeV data of fig.~2 compared with a Monte Carlo simulation evaluated at 
an excess energy of $Q=4.3$~MeV, where the fit leads to a value of
$\chi^2=0.8$. The variation of $\chi^2$ with $Q$ shown in (b) allows us to 
estimate the energy to be $Q=(4.3\pm 0.9)$~MeV.}
\end{figure}

\end{document}